\documentclass[pre,aps,twocolumn,floatfix]{revtex4}
\usepackage{graphicx}
\usepackage{dcolumn}
\usepackage[rgb,dvipsnames]{xcolor}
\usepackage{latexsym}
\usepackage{amsmath, amsthm, amssymb}
\usepackage{epsfig}
\usepackage{latexsym}
\usepackage{bm}

\begin{document}

\title{Optimal interdependence enhances robustness of complex systems}

\author{R. K. Singh and Sitabhra Sinha}
\affiliation{The Institute of Mathematical Sciences, CIT Campus, Taramani, Chennai 600113, India}

\begin{abstract}
%
%
While interdependent systems have usually been associated with increased
fragility, we show that strengthening the interdependence between dynamical
processes on different networks can make them more robust. By
coupling the dynamics of networks that in isolation exhibit
catastrophic collapse with extinction of nodal activity, we
demonstrate system-wide persistence of activity for an optimal range
of interdependence between the networks. This is related to the
appearance of attractors of the global dynamics comprising disjoint
sets (``islands'') of stable activity.
\end{abstract}

\pacs{PACS}

\maketitle

Many complex systems that occur in biological~\cite{Pocock2012},
technological~\cite{Rinaldi2001} and
socio-economic~\cite{Schweitzer2009} contexts
are strongly influenced by the behavior of
other systems~\cite{Helbing2013}.
Such interdependence can result in perturbations in one system
propagating to others, potentially resulting in a cascading
avalanche through the network of networks~\cite{vespignani,Gao2011}. 
Recent
studies of percolation of failure processes in a system of
two~\cite{Buldyrev2010,Parshani2010,Wei2012} or
more~\cite{Gao2012} connected networks have suggested that interdependence
makes the entire system fragile. However, a proper appraisal
of the role of interdependence on the stability of complex systems
necessarily needs to take into account the dynamical processes
occurring on them~\cite{Barzel2013,Danziger2015}. Compared to a purely
structural approach (such as percolation, that considers the effect of
removing nodes or links), a dynamical systems
perspective provides a richer framework for
assessing the robustness of systems.
Indeed, investigating how fluctuations from the equilibrium state in a local
region of a complex system can propagate to other regions forms the basis for
addressing the dynamical stability of the system~\cite{May1972}. Extending this
framework to the context of interdependent networks can
potentially offer us insights on why such systems are
ubiquitous in the real world in spite of their structural fragility (as
indicated by percolation).


In this paper we show that strong interdependence between networks can
{\em increase} the robustness of the
system in terms of its dynamical stability.
In particular, we show for a pair of
networks, whose nature of inter-network coupling differs
qualitatively from the intra-network links, that there exists an
optimal range of interdependence which substantially enhances the persistence probability of active nodes.
By contrast, decreasing
the inter-network coupling strength 
so that the networks are effectively independent
results in a catastrophic
collapse with extinction of activity in the system almost in its
entirety. The increased persistence at optimal coupling
is seen to be related to the appearance of
attractors of the global dynamics comprising disjoint sets
of stable activity.
Our results also suggest that
the nature of inter-network interactions is a crucial
determinant of the role of interdependence on the robustness of
complex systems. For example, increasing the intensity
of nonlinear interactions between nodes
leads to loss of stability and subsequent
transition of the nodes to a quiescent state, while
stronger diffusive coupling between the networks
can make a global state corresponding to persistent activity 
extremely robust.

Let us consider a model system comprising $G$ interdependent
networks. Each network has $N$ dynamical elements connected to each
other through a sparse random
topology of nonlinear interactions.
Interdependence is introduced by diffusively coupling an element $i$
in a network to the corresponding $i$-th element of other network(s).
This framework can be used to represent,
for instance, dispersal across $G$ neighboring habitat patches of $N$
interacting species
in an ecological system.
A continuous dynamical variable $z_i^{\mu}$ ($i=1,\ldots,N; \mu = 1,
\ldots,G$) is associated with each node of the multiple coupled
networks. In the above-mentioned example, it can be interpreted as the
relative mass density of the
$i$-th species in the $\mu$-th patch. We consider generalized
Lotka-Volterra interactions between the nodes within a
network as this is one of the simplest and
ubiquitous types of nonlinear coupling~\cite{Goel1971,sinha}. The dynamical
evolution of the system can then be described in terms of the $GN$ coupled
equations:
\begin{multline}
  \label{dynrule1}
  z_i^{\mu}(n+1) = (1 - D) F(z_i^{\mu}(n))[1+\sum_{j=1}^{N}
  J^{\mu}_{ij} z_j^{\mu}(n)] \\
  + \frac{D}{(G-1)} \sum^G_{\nu \neq \mu} ~F(z_i^{\nu}(n))[1+\sum_{j=1}^{N}
  J^{\nu}_{ij} z^{\nu}_j(n)].
\end{multline}
Here {\bf J}$^\mu$ is the interaction matrix for the $\mu$-th network, 
while $D$ is a measure of the strength of interdependence via
diffusive coupling between networks. 
The range of the variable $z_i^{\mu}$ is decided by the
function $F()$ governing the dynamics of individual elements in
the system. Here we consider $F$ to be a
smooth unimodal nonlinear map defined over a finite support and having an
absorbing state. This class of dynamical systems is quite general and
are capable of exhibiting a wide range of behavior including
equilibria, periodic oscillations and chaos~\cite{Feigenbaum1978}.
For the results shown here we have used the logistic
form~\cite{logisticmaps}: $F(z) = r z
(1-z)$ if $0<z<1$, and $=0$ otherwise, such that $z=0$ is the
absorbing state, and $r$ is a nonlinearity parameter that determines
the nature of the dynamics. Unlike most studies involving logistic map
that consider $r \in [0,4]$ so that the dynamics is confined to the unit
interval, here we specifically use $r > 4$. This implies that $F()$
maps a finite subinterval within $[0,1]$ 
directly to the absorbing state. Iterative application of $F()$ will result in
only a set of measure zero remaining in (0,1)~\cite{Brin2002}.
Thus, in isolation,
the nodes will almost always converge to the absorbing state, resulting in
their extinction.

However, interaction with other nodes can maintain
a node in an active state ($z>0$) indefinitely, and 
we define a measure for the global stability of the system as 
the asymptotic fraction of nodes in each network that have not
reached the absorbing state, viz., 
$f_{active} = {\rm Lt}_{n \rightarrow \infty} f_{active} (n)$, where
$f_{active}(n)=\sum_{i = 1}^{N} \Theta[F(z^\mu_i(n))]/N$ (with $\Theta[x] = 1 ~\text{for}~ x > 0, ~\text{and}~ 0
~\text{otherwise}$).
Thus, we explicitly investigate conditions under which interdependence
between networks can result in persistent activity in at least a
subset of the nodes comprising the system. Using an ecological
analogy, our focus is on
the long-term survival of a
finite fraction of the ecosystem as a function of the degree of
dispersal between neighboring patches
rather than the intrinsic stability of individual species
populations.

\begin{figure}
\includegraphics[width=0.5\textwidth]{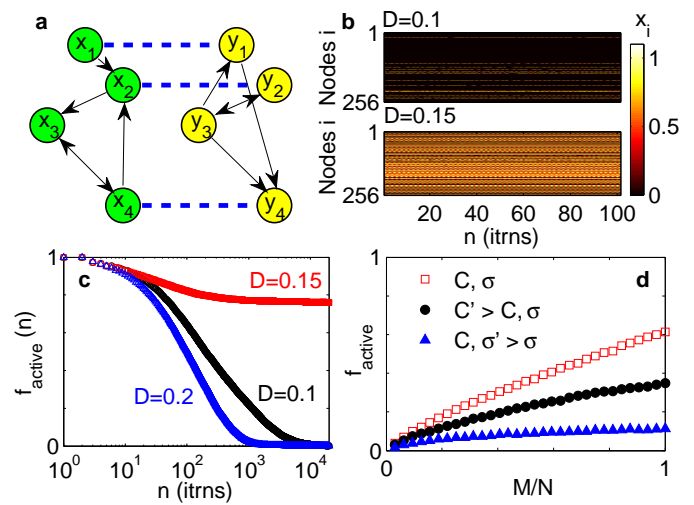}
\caption{
(a) Schematic diagram representing two interdependent
networks, each comprising $N$ nodes, that have intra-network 
directed nonlinear
interactions (indicated by arrows) and inter-network diffusive coupling
between $M$ ($\leq N$) pairs of corresponding nodes
(broken lines).
(b) Pseudocolor representation of the spatio-temporal evolution of 
dynamical state $x_i$ for each node $i$ in
one of the networks at two different values of the
inter-network diffusive coupling strength, viz., (top) $D = 0.1$ and
(bottom) $D = 0.15$, with black representing the absorbing state $x_i
=0$, i.e., extinction of activity. 
Increased interdependence between the networks allows
more nodes to maintain persistent activity, i.e., $x \neq 0$.
Increasing $D$ further can result in a decrease in
the fraction of active nodes $f_{active}$ with time as seen in (c), 
indicating that long-term persistent activity occurs only within an
optimal range of interdependence.
(d) Increasing intra-network interactions either in terms of the
connection density ($C$) or their strength ($\sigma$) for a given
inter-network diffusive coupling strength (e.g., $D=0.15$), 
results in a decrease in the fraction $f_{active}$ of nodes with persistent
activity. 
However, increasing the number of corresponding node pairs $M$ in the two
networks that are coupled diffusively
is seen to increase $f_{active}$, pointing to a fundamental
distinction between intra- and inter-network interactions in their
contribution to
the overall robustness of the system.
Parameter values for the curves shown are $C = 0.1, C' = 0.3, 
\sigma = 0.01, \sigma' = 0.05$ and $r\in [4.0,4.1]$.
Results shown here are obtained for $N=256$ and averaged over 100 ensembles. 
}
\label{timeseries}
\end{figure}

The degree of interdependence between the networks can be varied by
changing the number of pairs of corresponding nodes $M$ ($0 \leq M
\leq N$) that are linked via dispersion.
The interaction matrix {\bf J}$^{\mu}$ in each network is considered to be
sparse, such that only $C$ fraction of the matrix elements are non-zero
with their interaction strengths chosen randomly from a
$Normal(0,\sigma^2)$ distribution. 
For simplicity, we shall focus on a pair of interdependent networks
(i.e., $G = 2$) schematically shown in Fig.~\ref{timeseries}~(a), both
networks being chosen from the same ensemble so as to
have identical parameters $r$, $C$ and $\sigma$. We distinguish
between the population densities $z$ of the $N$ species in the two
networks by denoting them as $x_i$ and $y_i$
($i=1,\ldots,N$) respectively, their initial values being chosen at
random from the uniform distribution [0,1].

Fig.~\ref{timeseries}~(b-c) show the time-evolution of the state of
the dynamical variables $x_i$ and the global stability measure
$f_{active}(n)$ for one of the networks
($N=256$, $C = 0.1$, $\sigma =
0.01$) where the nonlinearity parameters $r_i$ are distributed uniformly in
[4.0,4.1].
As mentioned above, this distribution of $r_i$ implies that the
individual node dynamics would almost certainly converge to the
absorbing state, and this is indeed what is observed when
the networks are isolated, i.e., $D=0$. However, when the
interdependence is increased, e.g., to $D=0.15$, we observe that a
finite fraction of
nodes persist in the active state,
although for much lower (e.g., $D=0.1$) and
higher (e.g., $D=0.2$) interdependence the system exhibits complete 
extinction of activity [Fig.~\ref{timeseries}(b-c)].
Thus an optimal diffusive coupling between corresponding nodes in the two
networks provides global stability to the system. This suggests, for
instance, that
ecological niches which in isolation are vulnerable to systemic
collapse resulting in mass extinction, can retain species
diversity if connected to neighboring habitats through species
dispersal. Indeed, for this to happen, it is not even required that
all species in the network be capable of moving between the different
habitats. As seen from Fig.~\ref{timeseries}~(d), if only a subset of
$M$ nodes (out of $N$) are coupled between the two networks through
diffusion, the system exhibits enhanced persistence with $f_{active}$ 
increasing with $M$.
However, enhancing the intensity of nonlinear interactions within each
network by increasing either their connectivity $C$ or range of
interaction strengths (measured by the dispersion $\sigma$)
decreases the survival probability of active nodes,
as has been observed in earlier studies of
global stability of independent networks~\cite{Sinha05,sinha}.

\begin{figure}
\includegraphics[width=0.5\textwidth]{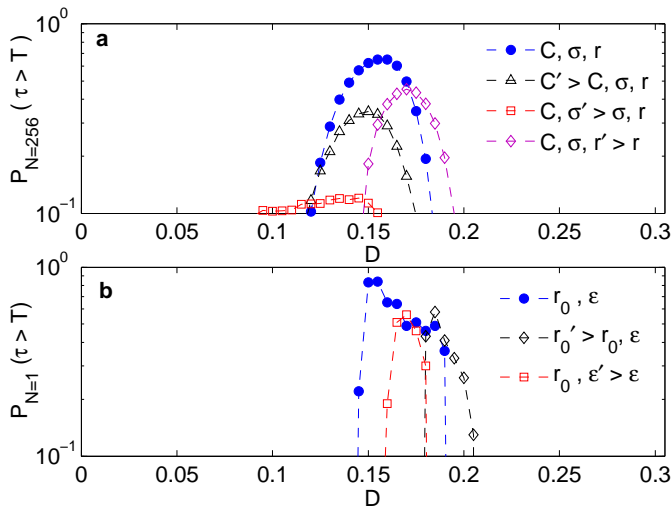}
\caption{
(a) Probability that nodal activity persists for longer than the duration of
simulation $P(\tau > T)$ for an interdependent system of two
networks has a non-monotonic
dependence on the inter-network coupling strength $D$ but decreases
with increasing $C$, $\sigma$ and $r$.
Each of the networks comprise $N=256$ nodes.
Parameter values used are $C = 0.1$, $\sigma =
0.01$, $r \in [4.0,4.1]$, $C^{\prime} = 0.3$, $\sigma^{\prime} = 0.05$,
and $r^{\prime} \in [4.1,4.2]$. 
(d) Probability of persistent activity in a system of two diffusively coupled
elements ($N=1$) whose nonlinearity parameters fluctuate about $r_0$
due to a noise of strength $\epsilon$.
Non-monotonic
dependence on coupling strength $D$ is seen, similar to that for the
large networks shown in (a).
Parameter values are $r_0 = 4.05$, $\epsilon = 0.005$, ${r_0}^{\prime}
= 4.2$ and $\epsilon^{\prime} = 0.01$.
For all panels, simulation duration is $T = 2 \times 10^4$ itrns and 
results shown are averaged over 100 ensembles.
}
\label{firstpassage}
\end{figure}


Fig.~\ref{firstpassage}~(a) shows in detail the contrasting
contribution of intra- and inter-network interactions to the
robustness of the network in terms of maintaining persistent activity.
The probability that a node persists in the active state asymptotically
is seen to vary non-monotonically with increasing interdependence $D$ between
the networks at different values of the
parameters $C$, $\sigma$ and $r$ that determine intra-network dynamics. 
For reference let us
focus on the curve for $C=0.1$, $\sigma = 0.01$ and $r \in [4.0,
4.1]$ [shown using circles in (a)]. We observe that when diffusion is
either too low ($D < 0.09$) or high ($D>0.2$) all activity in the network
ceases within the duration of simulation. However, for the
intermediate range of values of $D$, activity continues in at least a
part of the network with the persistence probability reaching a peak
around $D \simeq 0.16$.
Varying the other parameters, such as network connectivity $C$,
intra-network interaction strength $\sigma$ or the
nonlinearity parameter $r$, has a simpler outcome, viz., a 
decrease in the
probability that activity will persist in the network at long times.
This is shown by the other curves where we increase in turn $C$
(triangles),
$\sigma$ (squares) and $r$ (diamonds). Thus, our results indicate that
there exists an attractor
corresponding to persistent activity in the network
for an optimal range of interdependence (in the
neighborhood of $D=0.15$) which coexists with
the attractor corresponding to the extinction of network activity,
relatively independent of other parameters.

To understand this in detail, we first note that 
even when $N=1$, this much simpler system of two diffusively coupled
elements exhibits
qualitatively similar
features when subjected to noise [Fig.~\ref{firstpassage}~(b)]. 
The multiplicative noise of
strength $\epsilon$ in the nonlinearity parameter, viz., $r = r_0 (1 + \epsilon
\eta)$ where $\eta$ is a Gaussian
random process with zero mean and unit variance, is introduced in lieu
of the perturbations that each map will feel when connected to a much
larger network through nonlinear interactions
(Eq.~\ref{dynrule1})~\cite{note1}.
As in the case of the network, we choose $r_0 > 4$ so that an isolated
node will almost always converge to the absorbing state, resulting in
its extinction.
Upon coupling two nodes, however, 
we observe that the probability of long-term
survival of activity in the system becomes finite at an intermediate
range of diffusive
coupling strengths (around $D = 0.15$), 
similar to that observed for a $N=256$ network
in Fig.~\ref{firstpassage}~(a).
Thus, understanding the genesis of diffusion-induced persistence for
a pair of coupled logistic maps subject to noise~\cite{linz}, 
may provide an explanation for the same
phenomenon observed in the system of interdependent networks described
earlier.

\begin{figure}
\includegraphics[width=0.5\textwidth]{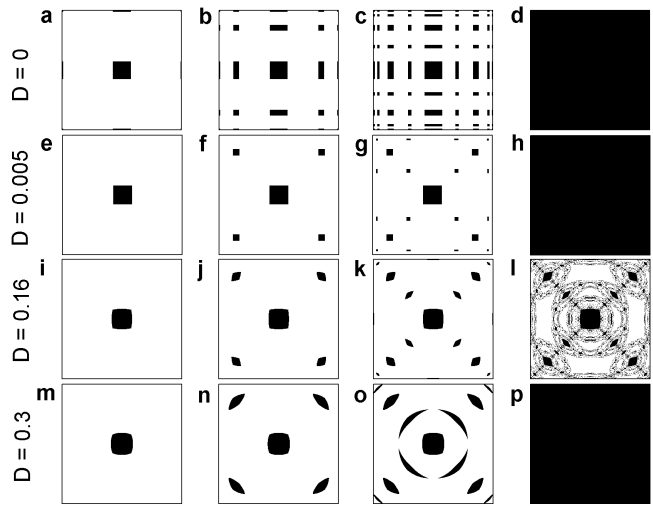}
\caption{Time-evolution of a system of two diffusively coupled
logistic maps having $r>4$, showing (in black) the regions of phase
space $I^2: [0,1]\times[0,1]$ that
correspond to initial states which lead to trajectories moving out of
the $I^2$ domain resulting in extinction of activity
after $n=1$ (1st column), $=2$ (2nd column), $=3$ (3rd column) and
$=2000$ iterations (4th column) in both the maps. 
As each map has a segment
projecting out of $I^2$, repeated iteration of the system
when the maps are isolated ($D=0$) would eventually drive almost all
initial states to extinction (a-d). The same behavior is also observed for
a low degree of diffusive coupling ($D=0.005$, e-h), although the 
regions are now modulated because of the coupling with the dynamics
of the other map. For a stronger diffusive coupling ($D=0.16$) 
there is a finite region of phase space that remains within $I^2$ even
after a large number of iterations which corresponds to the basin for
the attractor exhibiting persistent activity (i-j). Further increase in
the coupling strength (e.g., $D=0.3$) again results in extinction of
activity for almost the entire phase space (m-p).}
\label{basin}
\end{figure}

Fig.~\ref{basin} shows the regions in the phase space $I^2:
[0,1] \times [0,1]$ of the system of two coupled
maps that correspond to initial states which move out of $I^2$ and
into the absorbing state as a result of the dynamics.
When the maps are isolated ($D=0$), successive iterations
result in these regions increasing in size and eventually taking over
the entire domain so that extinction will always happen. Similar
behavior is seen for weak coupling ($D=0.02$) although the shape of
the regions are now modulated as a result of the interaction between
the two maps. For high values of coupling also ($D=0.3$), we observe the total
extinction of activity in the asymptotic limit. However, for an
intermediate value of coupling ($D=0.16$), the complement region
defined by trajectories starting from anywhere inside it remaining
within $I^2$, retains a finite measure even at long times, thereby ensuring
persistence of activity. Introducing multiplicative noise in the
dynamics
does not significantly change the structure of the basins 
shown in Fig.~\ref{basin} for low values of the noise 
strength $\epsilon$.


To understand the above results we represent the dynamics of the system as
$x_{n+1} = (1-D) F(x_n) + D F(y_n)$, $y_{n+1} = (1-D) F(y_n) + D F(x_n),$
where $x,y$ are dynamical variables and $F(x) = rx(1-x)$. The first
term of each evolution equation can be interpreted as
a logistic map with growth rate $(1-D)r$
while the second term represents a contribution from the other map
that is diffusively coupled to it. 
If the sum of the two terms exceed 1 for any of $x,y$,
the corresponding variable goes to the absorbing state. 
In the weak coupling limit of low $D$ where the first term dominates, it follows
that if the effective growth rate $(1-D)r$ exceeds 4, the system will
exit the unit interval almost surely.
Thus a lower bound for the range of $D$ in which persistence can be
observed is obtained by ensuring that $D > D_{c1} = 1-(4/r)$. For
example, for $r=4.1$, $D_{c1} \simeq 0.024$.
The upper bound of $D$ for persistence is obtained by
observing that when the two coupled maps synchronize their activity 
upon strong coupling, the dynamics
reduces to that of an effective 1-dimensional map with $r>4$ whose
trajectories will eventually exit the unit interval with probability 1.
Whether
synchronization occurs can be investigated by looking at the dynamics
of the difference of the two variables, $\delta = y - x$, viz., $\delta_{n+1}
= r (1-2D) \delta_n [1-(x_n+y_n)]$. If $D>D_{c2}=(1/2)(1-(1/r))$,
e.g., $D_{c2} \simeq 0.37$ for $r = 4.1$, the
difference goes to zero asymptotically resulting in synchronization of
the two maps and convergence to the absorbing state. Thus, the system has a
possibility of persistence only in the intermediate range $D_{c1} < D
< D_{c2}$. 
In this region, where the individual maps exhibit periodic attractors,
persistence can arise through out-of-phase oscillations in the two
maps, each alternately visiting two disjoint intervals in [0,1] 
such that the sum of terms never exceed 1.
Thus, regions of $[0,1]\times[0,1]$ domain which yield the in-phase solution
lead to the absorbing state (extinction), 
while those giving rise to the out-of-phase
solution lead to persistence, resulting in a complex basin of
attraction for the persistent activity state [shown in
Fig.~\ref{basin}~(i-l)]. To show that such stable out-of-phase period-2 
solutions exist for an optimal range of $D$, we can solve the coupled
equations
$x^*_{1,2} = (1-D) F(x^*_{2,1}) + D F(x^*_{1,2}),$
and check for stability, thereby obtaining an implicit equation
involving the parameters $r$ and $D$ for which $0<x^*_1,x^*_2<1$. For
specific choices of $r$ and $D$, we can numerically verify that
these solutions are stable, thereby providing confirmatory evidence of
the proposed mechanism by which an optimal coupling induces persistence.

\begin{figure}
\includegraphics[width=0.5\textwidth]{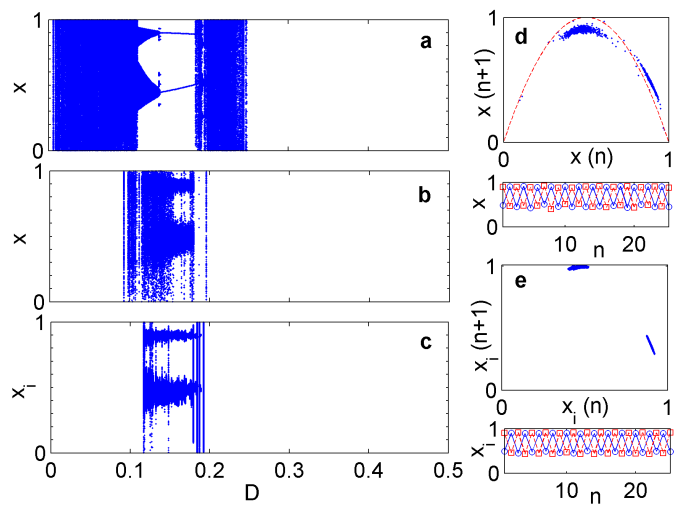}
\caption{Bifurcation diagrams showing the attractor of the
dynamical state $x$ of a representative element as a function of the
diffusive coupling strength between (a-b) two maps and (c) two
networks each comprising $N = 256$ nodes. (a-b) The range of $D$ over
which there is long-term persistence of activity in two coupled maps
for $r=4.0025$ (a) is reduced when multiplicative noise of strength
$\epsilon = 0.0125$ is introduced (b). The bifurcation structure
resembles that of coupled networks shown in (c) for 
$M = 24$, $C = 0.1, \sigma = 0.01, r \in [4.0, 4.1]$. The contribution
of intra-network interactions is qualitatively similar to
multiplicative noise, resulting in a similar range of $D$ for which
persistence is observed in (b) and (c). (d-e) This similarity is reinforced
by a comparison the return maps (upper panels) and time-series (lower
panels) of the asymptotic dynamical states for (d) two coupled maps
with noise [as in (b)] and (e) two networks [as in (c)] for
$D=0.15$.
The broken curve in panel (d) represents the return map of an
uncoupled logistic map for $r=4.0025$ shown for comparison.
}
\label{bifdgm}
\end{figure}
The bifurcation diagrams shown in Fig.~\ref{bifdgm}~(a-c) indicate how 
the range of diffusive coupling strengths over which persistent
activity is observed changes as we move from the simple case of
two coupled maps ($N=1$) to interdependent networks ($N >>1$). 
As already discussed, diffusively coupling two logistic maps having
$r>4$ allow their states to remain in the
unit interval (corresponding to the nodes being active) provided the strength
of coupling $D$ remains within an optimal range
[Fig.~\ref{bifdgm}(a)]. Note that within this range there exists a
region, approximately between (0.11,0.18), in which the attractor of the
dynamical state of the node occupies a much smaller region of the
available phase space $I$. It is intuitively clear that for such
values of $D$, introducing noise is much less likely to result in the
system dynamics going outside the unit interval (thereby making the
node inactive).
If we now introduce multiplicative noise of low intensity (i.e., small
$\epsilon$), the range of $D$ over 
which persistent activity occurs shrinks [Fig.~\ref{bifdgm}(b)].
However, noise does not completely alter the nature of the system dynamics even
though the the bifurcation structure is now less crisp.
In particular, the system appears to be particularly robust in the region
referred to earlier where the attractor covers only a
small volume of the unit interval. 
We can compare this case with that of two interdependent networks, each
comprising a large number of nodes [Fig.~\ref{bifdgm}(c)]
where the intra-network interactions are considered effectively to
be `noise'. 
We observe a reasonable similarity between their
bifurcation structures, with the region of persistent activity
spanning approximately the same range of $D$. As in the case of
coupled maps with noise, in the case of networks also the
system is most robust in the region where the attractor for the 
unperturbed system of two diffusively coupled maps is confined
within a small subinterval inside $I \times I$.
The validity of considering the dynamics of a coupled nodes embedded within a
network as equivalent to the pair being perturbed by an effective
external noise is further established by the strong resemblance
between the return maps and time-series for the two cases
[Fig.~\ref{bifdgm}(d-e)] for a diffusive coupling strength in the
optimal range that leads to persistent activity. 
As mentioned earlier, to survive indefinitely the dynamical state of
each map switches alternately
between two disjoint intervals of the unit interval in an out-of-phase
arrangement [see the time-series in the lower panels of
Fig.~\ref{bifdgm}(d-e)], corresponding to a
trajectory that jumps between two
``islands'' of the basin for the attractor corresponding to persistent activity attractor of the coupled
system shown in
Fig.~\ref{basin}~(l).


The above analysis, apart from explaining why populations that go
extinct rapidly in isolation will survive for long times upon being
coupled optimally, also helps us understand how the
persistence behavior in the system will be affected by increasing the
number of interacting components.
As can be observed from Eq.~(\ref{dynrule1}),
increasing $N$ keeping $C$, $\sigma$ unchanged corresponds to the
summation in the interaction term being performed over more
components. This suggests that there will be stronger
fluctuations, that can be interpreted as a larger effective
noise applied to the individual elements resulting in a higher probability
of reaching the absorbing state and thereby lowering the 
survival fraction $f_{active}$. We have confirmed this through
explicit numerical calculations in which $N$ is systematically
increased. To ensure that the
results reported here are not sensitively dependent on the specific
details of the model that we have considered here, we have also
carried out simulations with (a) 
different forms of  unimodal nonlinear maps, e.g., 
$F(x) = (x-l) e^{r(1-x)}~\text{for}~ x>l;~0
~\text{otherwise}$~\cite{Ricker1954}, and (b)
different types of connection topologies for the initial network,
e.g., those with small-world properties~\cite{Watts1998} or having
scale-free degree distribution~\cite{Barabasi99}.
We find in all such cases that that the qualitative features reported here are 
unchanged, with the network connecting the surviving nodes becoming
homogeneous asymptotically
irrespective of the nature of the initial topology, suggesting that
the enhanced persistence of activity in optimally interdependent networks
is a generic property.

To conclude, we have investigated the role of interdependence between
constituent networks on the stability of the entire system in a
dynamical framework. Unlike percolation-based approaches where failure
is often identified exclusively with breakdown of
connectivity so that increasing interdependence necessarily enhances
fragility~\cite{vespignani}, 
our dynamical perspective that 
considers processes on nodes
with a diverse spectrum spanning equilibrium, periodic and chaotic
behavior, reaches a strikingly different conclusion. In
particular, we show that if the physical interdependence between two
networks has the form of diffusive coupling between corresponding
nodes, the system has a much higher likelihood of survival when the
coupling strength is in an optimal range, even if, in isolation both
networks face almost certain catastrophic collapse. Thus,
interdependence between complex networks need not
always have negative repercussions. Instead its impact depends
strongly on the context (e.g., the nature of the interdependence
and the type of dynamics being considered).
It could well be the case that there are several real systems, e.g.,
in ecology, where interdependence is essential for maintaining
diversity in the presence of persistent fluctuations that
are potentially destabilizing. Instead of a simple equation between
interdependence and fragility, future investigations into systems
exhibiting diverse dynamics and variety of possible
couplings may reveal a more nuanced picture of the benefits
and drawbacks of interdependence and the trade-offs
involved in specific settings.

We would like to thank Trilochan Bagarti, Abhijit Chakraborty, Deepak Dhar, V. Sasidevan
and Amit Sharma for helpful discussions.


\begin{thebibliography}{10}
\bibitem{Pocock2012}
M.~J.~Pocock, D.~M.~Evans and J.~Memmott,
Science {\bf 335}, 973 (2012).

\bibitem{Rinaldi2001} S.~M.~Rinaldi, J.~P.~Peerenboom and T.~K.~Kelly, IEEE Control Syst. {\bf 21}, 11 (2001).

\bibitem{Schweitzer2009}
F.~Schweitzer, G.~Fagiolo, D.~Sornette, F.~Vega-Redondo,
A.~Vespignani and D.~R.~White,
Science {\bf 325}, 422 (2009).

\bibitem{Helbing2013}
D.~Helbing, Nature (Lond.) {\bf 497}, 51 (2013).

\bibitem{vespignani} A.~Vespignani, Nature (Lond.) {\bf 464}, 984 (2010).

\bibitem{Gao2011} 
J.~Gao, S.~V.~Buldyrev, S.~Havlin and H.~E.~Stanley,
Phys. Rev. Lett. {\bf 107}, 195701 (2011).

\bibitem{Buldyrev2010} 
S.~V.~Buldyrev, R.~Parshani, G.~Paul, H.~E.~Stanley and S.~Havlin,
Nature (Lond.) {\bf 464}, 1025 (2010).

\bibitem{Parshani2010}
R.~Parshani, S.~V.~Buldyrev and S.~Havlin, 
Phys. Rev. Lett. {\bf 105}, 048701 (2010).

\bibitem{Wei2012} W.~Li, A.~Bashan, S.~V.~Buldyrev, H.~E.~Stanley and
S.~Havlin, Phys. Rev. Lett. {\bf 108}, 228702 (2012). 

\bibitem{Gao2012} J.~Gao, S.~V.~Buldyrev, H.~E.~Stanley and S.~ Havlin, 
Nature Phys. {\bf 8}, 40 (2012). 

\bibitem{Barzel2013} B.~Barzel and A.-L.~Barab\'{a}si, Nature Phys. {\bf
9}, 673 (2013).
\bibitem{Danziger2015}
M.~M.~Danziger, A.~Bashan and S.~Havlin,
New J. Phys. {\bf 17}, 043046 (2015).
\bibitem{May1972}
R.~M.~May, Nature (Lond.) {\bf 238}, 413 (1972).
\bibitem{Goel1971} N.~S. Goel, S.~C. Maitra and E.~W. Montroll,
Rev. Mod. Phys., {\bf 43}, 231 (1971).
\bibitem{sinha} S. Sinha and S. Sinha, Phys. Rev. E. {\bf 74}, 066117 (2006).
\bibitem{Feigenbaum1978} M.~J.~Feigenbaum, J. Stat. Phys., {\bf 19}, 25 (1978). 
\bibitem{logisticmaps} R.~M.~May, Nature, {\bf 261}, 459 (1976). 
\bibitem{Brin2002}
M.~Brin and G.~Stuck,
{\it Introduction to Dynamical Systems} (Cambridge Univ. Press, Cambridge,
2002).
\bibitem{Sinha05} S. Sinha and S. Sinha, Phys. Rev. E. {\bf 71},
020902 (2005).
\bibitem{note1}
The choice of Gaussian noise is 
motivated by the strength of intra-network interactions, $J_{ij}$, 
being drawn from a normal distribution of zero mean and a finite
variance. Moreover, if we focus on a single time-evolution step, the
perturbations arising from intra-network interactions can be
approximated by Gaussian distributed random variables for large
networks by using the Central Limit Theorem.
Subsequent evolution of the network may introduce
correlations among the perturbations arising from the interactions
which will not be accounted for in this Gaussian noise approximation.
\bibitem{linz} S. J. Linz and M. Lucke, Phys. Rev. A {\bf 33}, 2694 (1986). 
\bibitem{Ricker1954} Similar in form to the population dynamics model
proposed in W.~E.~Ricker, J. Fish. Res. Board Can. {\bf 11},
559 (1954).
\bibitem{Watts1998} D.~J. Watts and S.~H. Strogatz, Nature (Lond.) {\bf
393}, 440 (1998).
\bibitem{Barabasi99} A.-L. Barab\'{a}si and R. Albert, Science {\bf
286}, 509 (1999).
\end{thebibliography}
\end{document}